\documentclass[review]{elsarticle}
\usepackage{booktabs}% http://ctan.org/pkg/booktabs

\usepackage{geometry}
\usepackage{array}
\usepackage{pdflscape}
\usepackage{multicol} % to merge columns
\usepackage{multirow} % to allow multiple rows in one cell
\usepackage{siunitx} % line up decimal points in tables
\usepackage[T1]{fontenc} % for latin characters
\usepackage[latin1]{inputenc} % for latin characters
\usepackage{textcomp} % for a 'TM' mark
\usepackage{epstopdf}
\journal{Radiotherapy \& Oncology}

\begin{document}

\begin{frontmatter}

\title{Modeling of Radiation Pneumonitis after Lung Stereotactic Body Radiotherapy: A Bayesian Network Approach}

\author[add1]{Sangkyu Lee}
\author[add1]{Norma Ybarra}
\author[add1]{Krishinima Jeyaseelan}
\author[add2]{Sergio Faria}
\author[add2]{Neil Kopek}
\author[add2]{Pascale Brisebois}
\author[add3]{Toni Vu}
\author[add3]{Edith Filion}
\author[add3]{Marie-Pierre Campeau}
\author[add3]{Louise Lambert}
\author[add3]{Pierre Del Vecchio}
\author[add3]{Diane Trudel}
\author[add3]{Nidale El-Sokhn}
\author[add4]{Michael Roach}
\author[add4]{Clifford Robinson}
\author[add1,add5]{Issam El Naqa}

%\cortext[cor1]{Please address correspondence to Author2 or Author3}
\address[add1]{Medical Physics Unit, McGill University, Montreal, Canada}
\address[add2]{Department of Radiation Oncology, Cedars Cancer Centre, Montreal, Canada}
\address[add3]{Department of Radiation Oncology, Centre Hospitalier de l'Universit\'e de Montr\'eal, Montreal, Canada}
\address[add4]{Department of Radiation Oncology, Washington University in St. Louis, St. Louis, United States}
\address[add5]{Department of Radiation Oncology, University of Michigan, Ann Arbor, United States}

\begin{abstract}
\noindent Background and Purpose: Stereotactic body radiotherapy (SBRT) for lung cancer accompanies a non-negligible risk of radiation pneumonitis (RP). This study presents a Bayesian network (BN) model that connects biological, dosimetric, and clinical RP risk factors.

\noindent Material and Methods: 43 non-small-cell lung cancer patients treated with SBRT with 5 fractions or less were studied. Candidate RP risk factors included dose-volume parameters, previously reported clinical RP factors, 6 protein biomarkers at baseline and 6 weeks post-treatment. A BN ensemble model was built from a subset of the variables in a training cohort (N=32), and further tested in an independent validation cohort (N=11).

\noindent Results: Key factors identified in the BN ensemble for predicting RP risk were ipsilateral V5, lung volume receiving more than 105\% of prescription, and decrease in angiotensin converting enzyme (ACE) from baseline to 6 weeks. External validation of the BN ensemble model yielded an area under the curve of 0.8.

\noindent Conclusions: The BN model identified potential key players in SBRT-induced RP such as high dose spillage in lung and changes in ACE expression levels. Predictive potential of the model is promising due to its probabilistic characteristics.

\end{abstract}


\end{frontmatter}

%\linenumbers

\section{Introduction}
In recent years, stereotactic body radiotherapy (SBRT) has become the treatment of choice for non-operable early stage non-small cell lung cancer (NSCLC), demonstrating a local control rate close to 90\% \cite{Timm10}. Incidence of pulmonary toxicity, usually defined as symptomatic radiation pneumonitis (RP), is reported to be less than 10\% \cite{Chi10} due to focused radiation to a small target which spares large volume of healthy lung \cite{Yamashita14}. However, several non-dosimetric factors reportedly increase or decrease the RP risk, such as central tumor location \cite{Timmerman06}, baseline interstitial pneumonitis \cite{Yamashita10} and chronic obstructive pulmonary disease (COPD) \cite{Takeda12}. Ignoring these factors could underestimate RP risk for certain patients. Thus, there is a clinical need to augment dosimetric RP models with patient-specific clinical and biological risk modifiers towards more patient-specific predictions.

We propose Bayesian network (BN) as a multivariate modeling platform to accommodate such high-dimensional data. BN can be characterized as graphical representation of relationships between input variables called a directed acyclic graph (DAG). Variables in a DAG are connected along the direction of influence. This allows us to study such a radiobiological system as a "whole" whereas conventional multivariate models such as logistic regression are limited to the predictive value of variables in a model \cite{IEN06}. The BN approach has been adopted by a number of outcome studies \cite{Smith09} \cite{Jayasurya10} \cite{Oh11b}, and specifically for radiation pneumonitis from conventional fractionation \cite{Lee15} where finding a consensus of prediction results from several BN models (ensemble approach) was shown to improve RP prediction. However, the BN approach has not been applied to SBRT cases where dose-volume metrics and biological damage relationships are still not well understood.

The aim of this study is to develop a Bayesian Network RP model for NSCLC SBRT patients. While a primary objective is to assess its predictive potential, we will also address its ability to uncover underlying radiobiological relationships and generate new hypotheses.

\section{Materials and Methods}
\subsection{Patient cohort}
Forty three stage I and II NSCLC patients were recruited for this study prospectively from three institutions upon approval of respective ethics review boards: McGill University Health Centre (MUHC), Centre Hospitalier de l'Universit\'e de Montr\'eal (CHUM), and Washington University in St. Louis (WashU), 32 patients from MUHC and CHUM formed the training cohort for BN modeling. 11 patients from WashU were reserved for model validation. Every patient met the following eligibility criteria: 1) received SBRT of equal or less than 5 fractions with curative intent, 2) no history of previous lung irradiation, and 3) baseline Karnofsky performance status $\geq$ 70. Detailed cohort characteristics are summarized in table \ref{tab:patients}. The patients were treated with radiotherapy (RT) without any adjuvant treatment. Depending on institutions, three different delivery techniques were used: 3-dimensional (3D) conformal radiotherapy, RapidArc\texttrademark (Varian Medical Systems, Palo Alto, CA) Volumetric Arc Therapy (VMAT), and CyberKnife (Accuray Inc, Sunnyvale, CA). Detailed RT procedures are provided in supplementary tables 1 and 2.

\begin{table}
\begin{center}
\caption{Characteristics of the training and validation cohorts. *Calculated for whole lung subtracted from planning target volume and converted to equivalent dose in 2 Gy fraction (EQD2).}
\begin{tabular}{ l c c  }
	\hline
	& \multicolumn{2}{c}{Patient count (\%)} \\ 
	& Training & Validation \\ \hline
	Cohort size & 32 & 11 \\
	Tumor stage & & \\
	\quad I & 32 (100) & 9 (81) \\ 
	\quad II & 0 (0) & 2 (19) \\
	RP grades & & \\
	\quad 0 & 17 (53) & 0 (0) \\
	\quad 1 & 11 (34) & 10 (91) \\
	\quad 2 & 2 (6) & 0 (0) \\
	\quad 3 & 2 (6) & 1 (9) \\
	\quad $\geq4$ & 0 (0) & 0 (0) \\
	\quad $\geq2$ & 4 (13) & 1 (9) \\
	Mean lung dose* & & \\
	\quad median & 4.9 & 6.3 \\
	\quad range & 2.4-10.9 & 1.2-9.9 \\
	RT modality & \\
	\quad 3D conformal & 19 (59) & 11 (100) \\
	\quad VMAT & 5 (16) & 0 (0) \\
	\quad CyberKnife & 8 (25) & 0 (0)  \\
	RT prescription & \\
	\quad 60 Gy in 3 fractions & 8 (25) & 0 (0) \\
	\quad 60 Gy in 5 fractions & 5 (16) & 1 (9) \\
	\quad 50 Gy in 5 fractions & 4 (13) & 4 (36) \\
	\quad 48 Gy in 3 fractions & 12 (38) & 0 (0) \\
	\quad 34 Gy in 1 fractions & 3 (9) & 0 (0) \\
	\quad 54 Gy in 3 fractions & 0 (0) & 5 (45) \\
	\quad 55 Gy in 5 fractions & 0 (0) & 1 (9) \\
\end{tabular}
\label{tab:patients}
\end{center}
\end{table}

\subsection{Data collection}
Blood samples from the patients were first acquired on the CT simulation day as a baseline and at 6 weeks post-treatment. Enzyme-linked immunosorbent assay (ELISA) was used for measuring biomarker concentrations in the samples. Incidence rate of symptomatic RP, classified as Common Toxicity Criteria for adverse events (CTCAE) toxicity (version 4) grade 2 or higher, was 13\% (4/32) in the training and 9\% (1/11) in the validation cohort. Median follow-up was 12 and 34 months for the training and validation cohorts, respectively. 

\subsection{Candidate variables}
Candidate variables for the BN pneumonitis model were chosen from 3 main categories: biological, dosimetric and clinical variables. Candidate biological variables consisted of serum concentration of interleukin(IL)-6, IL-8, angiotensin converting enzyme (ACE), alpha-2-macroglobulin ($\alpha$2M) , and transforming growth factor (TGF)-$\beta$1 and plasma concentration of osteopontin (OPN). As summarized in \cite{Fleckenstein07}, these biomarkers represent different biological processes involved in pathogenesis of radiation-induced lung injury, such as pro-(IL-6 \cite{Chen06}, OPN \cite{Oregan03}) and anti-(IL-8 \cite{Hart05}) inflammatory reactions, fibrogenesis (TGF$\beta$ \cite{Zhao07}), vascular damage (ACE \cite{Burnat91}) and modulation of inflammatory reactions ($\alpha$2M \cite{Oh11}). 12 features in total were extracted (6 biomarkers x 2 time points). The biomarker features at 6-weeks were calculated as percentage difference from respective baseline levels. The following 7 clinical RP risk factors were chosen by literature survey: superoinferior PTV location (PTVCOMSI) \cite{Hope06}, age \cite{Vogelius12}, smoking status \cite{Jin09}, COPD \cite{Takeda12}, ACE inhibitor \cite{Bracci15}, baseline interstitial lung disease \cite{Yamashita10}, and centrally located tumours \cite{Timmerman06}. Dosimetric factors were derived from planned dose converted to equivalent dose in 2 Gy fraction (EQD2) using an alpha-beta ratio of 4 Gy for lung \cite{Bentzen00} and 2 Gy for heart \cite{SchultzHector07}. For lung dose calculation, PTV was subtracted from contoured lung. Mean lung dose (MLD) and various Vx values (lung volume receiving > x Gy) for ipsilateral and whole lung were estimated. Due to high correlation between these parameters \cite{Marks10}, exploratory analysis was performed to find the smallest number of features that can capture dose heterogeneity relevant to RP. In this analysis, Vx was computed at various threshold dose x in three different ways: 1) x as an absolute dose or relative to prescription dose, 2) Vx normalized to lung volume or as a absolute volume, and 3) ipsilateral or whole lung. In addition to lung dose, we also considered mean heart dose (MHD) \cite{Huang11}, fraction size \cite{Roach95}, and PTV volume \cite{Allibhai13}.

\subsection{Bayesian Network training}
A Bayesian network ensemble model was trained from the candidate variables following the methods developed in \cite{Lee15}. In brief, the training was done in 4 steps:
\begin{enumerate}
\item Data discretization: Every continuous variable was discretized into 2 bins at a boundary that maximizes mutual information with respect to RP, as shown in supplementary table 3.
\item Feature selection with the Koller-Sahami (KS) filter: The number of candidate variables were reduced to the smallest subset that maximized explanatory power measured by cross-entropy with respect to RP. 
\item DAG training: Posterior distribution of Bayesian network graphs was obtained by Markov-Chain Monte Carlo sampling under causality constraints between variables. 
\item Parameter learning: Every variable in a BN is treated as a probabilistic distribution which is conditioned upon its upstream variables ("parents"). BN parameters, referred to as conditional probability values for every pair of a node and its parents, were learned from data using the expectation-maximization algorithm.
\end{enumerate}

\subsection{BN model testing}
The trained BN ensemble model derived probability of RP using known input variables. Classification of RP events was made by thresholding the RP probabilities. Classification performance was measured using three receiver operating characteristics (ROC) metrics: area under the curve (AUC), sensitivity, and specificity at the optimal operating threshold maximizing the sum of sensitivity and specificity. Model statistical testing was carried out in two ways: 1) the .632+ bootstrap method \cite{Efron97}: the training was repeated in 200 replicates which were resampled from the original data with replacement where the instances that were not sampled into the replicates were used for testing. 2) the final model was trained using the entire training set and tested in the validation cohort.

%model training was repeated in 200 bootstrap replicates which was resampled from the original data with replacement, and the instances that were not sampled into the replicates were used for testing. 

%\begin{figure}
%\centering
%\rotatebox{0}
%{
 % \scalebox{0.22}
 % {
 %   \includegraphics{SBRT_DVHs.eps} 
 % }
%}
%\caption{Superimposed DVHs from all the patients with (red curves) and without (blue curves) pneumonitis. Dose is normalized to 2-Gy per fraction equivalent (EQD2). Left: cumulative DVH with dose normalized to respective hot spot dose (107\% of a prescription dose) marked by a vertical line. Right: cumulative DVH with a raw EQD2 scale. The 5 Gy threshold is indicated in a vertical line.}
%\label{fig:binning}
%\end{figure}

\section{Results}
\subsection{Exploratory analysis of lung DVH parameters}
\begin{figure}
\centering
\rotatebox{0}
{
  \scalebox{0.36}
  {
  \hspace*{-6cm}
    \includegraphics{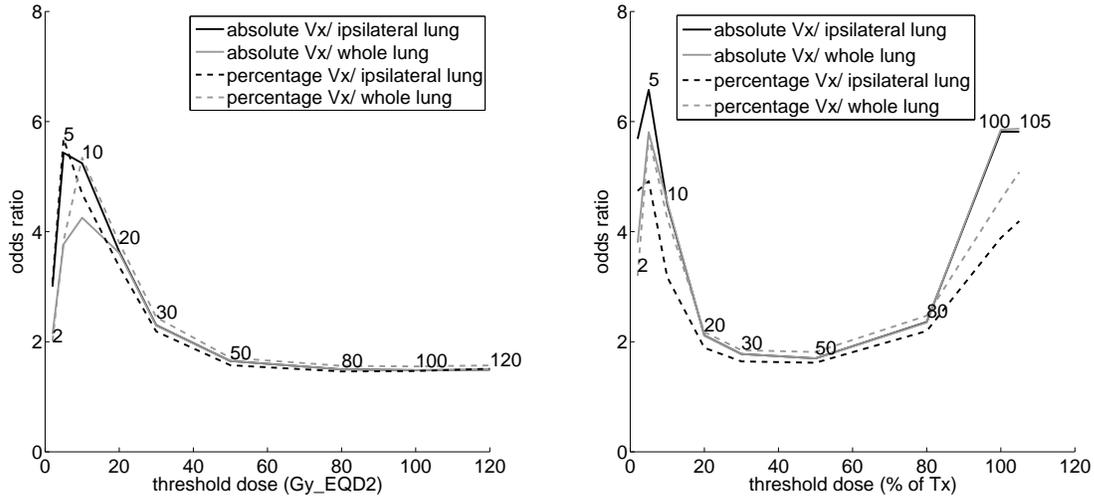} 
  }
}
\caption{Odds ratios of lung Vx measured at various threshold dose values (x), normalization schemes, and lung volume definition.}
\label{fig:Vx}
\end{figure}

Correlation between lung Vx and RP was examined by the change in odds ratios at various threshold dose (x) (figure \ref{fig:Vx}). When x was used as absolute dose, highest odds ratio (5.685) was marked at 5 Gy for ipsilateral lung. When the percentage of a prescription dose was used as x, increase in correlation was observed in high dose regions beyond 50\% of the prescription dose. Guided by this analysis, we chose two Vx parameters that represent low and high dose spillage in this order: percentage of ipsilateral lung volume receiving 5 Gy or more (V5) and absolute lung volume receiving more than 105\% of prescription dose (V105\%). In a similar fashion, ipsilateral MLD (odds ratio: 2.400) was preferred over MLD for the whole lung (odds ratio: 2.365).
  
\subsection{Variable selection and the Bayesian Network ensemble model}
The KS filter reduced the candidate (dosimetric, biological, and clinical) variables from 25 into the following 6 (supplementary table 3): 1) pre-treatment OPN, 2) 6 weeks ACE, 3) pre-treatment TGF$\beta$, 4) ipsilateral V5, 5) V105\%, and 6) PTVCOMSI. Inter-relationships between these variables and RP were established in the BN graphs. Bootstrap testing on BN graph learning detected 11 statistically significant links out of possible 19 from an ensemble of 50 graphs where bootstrapped RP prediction performance achieved optimality (figure \ref{fig:ROCmetricsBN}). A mean confidence level of the significant links was 0.57, while the upper bound of random variation was estimated to be 0.29 \cite{Scutari13}.

\begin{figure}
\centering
\rotatebox{0}
{
  \scalebox{0.4}
  {
  \hspace*{-6cm}
    \includegraphics{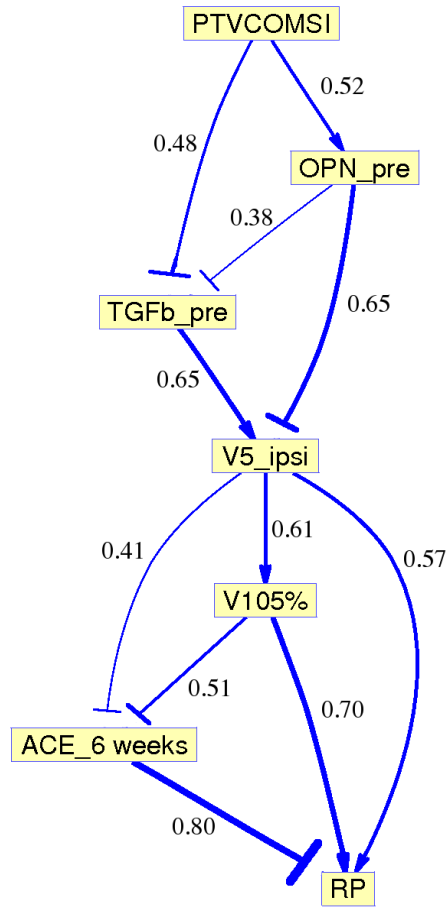} 
  }
}
\caption{Variables connected by significant associations detected in an ensemble of 50 graphs. Edge thickness was drawn proportionally to bootstrap estimated confidence level. Arrow-headed and bar-headed edges are assigned to positive and negative correlations, respectively. ipsi: ipsilateral lung, \_pre: baseline biomarker levels}
\label{fig:edgeconfidence}
\end{figure}

\subsection{Prediction performance of the BN model}
\begin{figure}
\centering
\rotatebox{0}
{
  \scalebox{0.31}
  {
  \hspace*{-6cm}
    \includegraphics{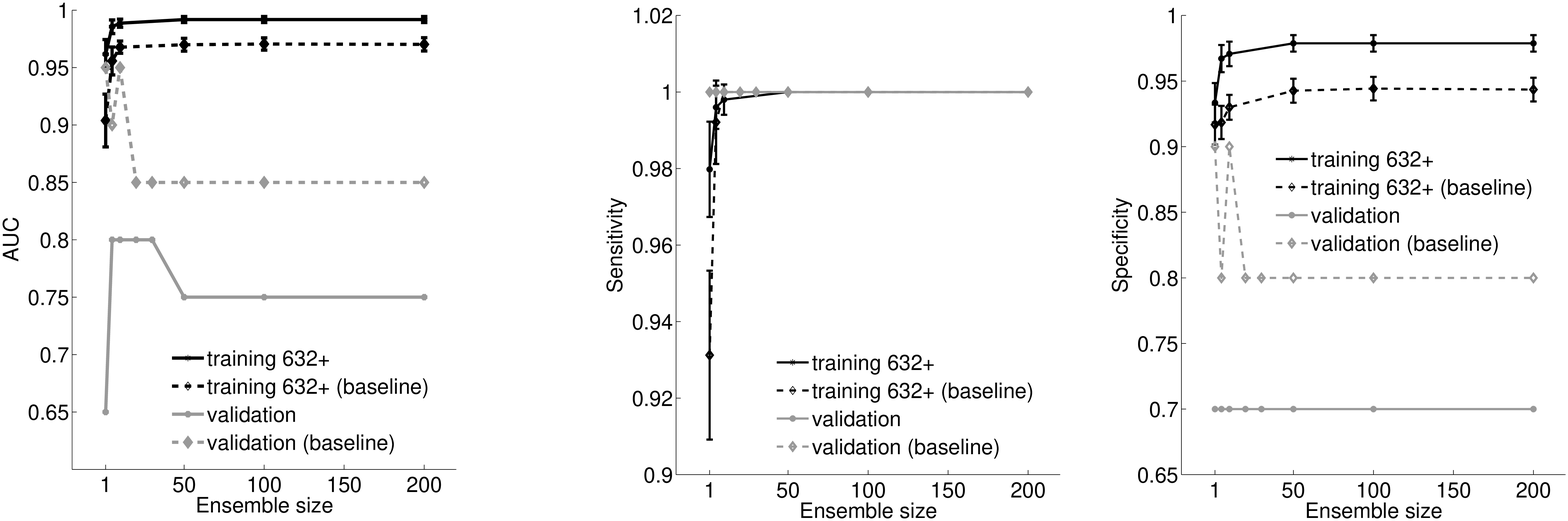} 
  }
}
\caption{Classification performance of the Bayesian Network model in two cohorts with varying ensemble size. Error bars: bootstrap-estimated 95\% confidence intervals. }
\label{fig:ROCmetricsBN}
\end{figure}

When bootstrap validation was used, RP prediction improved upon increasing number of graphs in an ensemble (figure \ref{fig:ROCmetricsBN}). Optimal performance was achieved at an ensemble size of 50 where AUC, sensitivity and specificity were 0.99, 1, and 0.98, respectively. At the optimal classification threshold, sensitivity was consistently higher than specificity. In external validation, AUC was the highest (0.8) at ensemble sizes 5-30. The BN model was subsequently tested using only the information available at baseline i.e. without ACE at 6 weeks. As a result, AUC and sensitivity decreased significantly at all ensemble sizes. In the validation cohort, however, slightly better performance was observed with only baseline information.

\section{Discussion}

Events of RP from lung SBRT are rare and identifying the susceptible patients before radiotherapy remains a difficult task, with conflicting results between studies. This study intended to objectively select and combine RP risk factors into a Bayesian Network and test its predictive power. Two factors account for good bootstrap performance of the resulting model in the training cohort. First, the main driving force was strong individual predictive power of the key variables in the model. Univariate AUC values of ACE at 6 weeks, V5, and V105\%, 3 variables connected to RP with high confidence, were respectively 0.94, 0.85 and 0.96 in the training cohort. Another factor was the use of an ensemble instead of a single model, which improved performance both in training and validation cohorts. Predictive benefit of an ensemble approach was already shown by other outcome studies \cite{Lee15} \cite{Das08}.

Amongst the biomarkers we studied, we found lower concentration of ACE at 6 weeks was strongly associated with RP events. This result is in line with investigation by Zhao et al. \cite{Zhao07} who reported lower ACE level at baseline and mid-treatment for patients with RP grade $\geq$ 2. In the Bayesian network ensemble the ACE was connected to dosimetric variables with high confidence. Causality of this relationship could be inferred from the knowledge that the main secretion site of ACE happens in lung epithelium and external stress to pulmonary vasculature such as ionizing radiation or Bleomycin exposure decreases serum ACE \cite{Burnat91}. The use of ACE inhibitors at baseline, however, was not a significant predictor of RP (p = 0.64) indicating that direct measurement of ACE expression could be a more sensitive test to predict RP.

Choice of 6 weeks as a time point to gauge post-treatment biomarker response was adequate to predict late toxicity before it happened, as the earliest occurrence of RP was 94 days post-RT. However, such information would not be available for the treatment planning stage inn such a case. We tested this scenario by attempting prediction with masked ACE values where the BN model can cope with missing information by marginalizing probability distribution over unknown variables. The role of ACE was not clearly shown in our external validation cohort where the absence of  such information did not reduce performance. Although these issues need to be confirmed in larger trial, our model predictions with post-RT variables may aid in monitoring high-risk patients or in prescribing anti-inflammatory medications.

We also observed that the size of high dose spillage, represented by lung volume outside PTV receiving dose > 105\%, was predictive of RP in univariate analysis and also one of the key variables in the BN model. This "high dose effect" on RP has been previously reported by a number of studies \cite{Hope06} \cite{Yamashita07}. Our results on exploratory analysis on Vx point out that both low-dose (V5) and high dose components might be relevant to RP. Previous lung SBRT protocols including RTOG 0236 and 0813 stipulate this volume as one of the quality assurance metrics to be regulated, setting its upper limit on 15\% of the PTV volume. Further studies may be needed to clarify the effects of smaller volume of high dose irradiation to lung leading to RP onset.

The main limitation of the current study is the low number of toxicity events in the evaluated cohorts, which led to relatively low specificity of the optimized model. Nevertheless, our computational approach reduced the data dimensionality and identified key variables that may mitigate the impact of a low event rate on fitting. Also, the links that we discovered in the BN graphs represent influential effects among variables that is not necessarily causal always. However, such knowledge may help provide new insights and guide generating new data-driven hypotheses.

%More accurate dose distribution could further improve the consistency of this model, considering that there are 3 different delivery techniques. We noticed that CyberKnife plans (N=8) were calculated with Monte Carlo, while convolution/superposition based algorithm was used for the rest. In order to test its impact on the model, we perturbed the values of dose bins of DVHs from the CyberKnife plans by 3\%, an estimate of difference between the two algorithms in lung \cite{Gagne07}. The resulting change in DVH parameters was not large enough to affect the prediction  

\section{Conclusion}
We developed a Bayesian Network ensemble for modeling radiation pneumonitis after lung SBRT. The process of building the model and the resulting model structure identified potential key players in predicting RP in NSCLC SBRT patients such as high dose spillage to the lung and changes in post-treatment ACE expression levels. This probabilistic model can potentially provide new insights into RP onset and help guide designing new studies as the interest of expanding SBRT to higher risk populations continues to grow. 

\section*{Conflict of Interest}
None.

\section*{Acknowledgement}
We thank Dr. Jean-Fran{\c c}ois Carri\`er and Dr. Robert Doucet for contributing to the clinical data. The computational work was enabled in part by computer resources provided by WestGrid (www.westgrid.ca). This research was partly funded by the Canadian Institute of Health Research (CIHR) grant MOP-114910. S.L  is supported by NSERC CREATE Medical Physics Research Training Network grant 432290.

\section*{References}
\bibliography{green2015_forarxiv.bbl}

%%%%%%%%%% Merge with supplemental materials %%%%%%%%%%
\pagebreak
\begin{center}
\textbf{\large Supplemental Materials for: Modeling of Radiation Pneumonitis after Lung Stereotactic Body Radiotherapy: A Bayesian Network Approach}
\end{center}
%%%%%%%%%% Merge with supplemental materials %%%%%%%%%%
%%%%%%%%%% Prefix a "S" to all equations, figures, tables and reset the counter %%%%%%%%%%
\setcounter{equation}{0}
\setcounter{figure}{0}
\setcounter{table}{0}
\setcounter{page}{1}
\makeatletter
\renewcommand{\theequation}{S\arabic{equation}}
\renewcommand{\thefigure}{S\arabic{figure}}
\renewcommand{\bibnumfmt}[1]{[S#1]}
\renewcommand{\citenumfont}[1]{S#1}
%%%%%%%%%% Prefix a "S" to all equations, figures, tables and reset the counter %%%%%%%%%%

\newgeometry{margin=2cm} % modify this if you need even more space
\begin{landscape}
\begin{table}
\begin{center}
\caption{Detailed radiotherapy procedures used for the training cohort. GTV: gross tumour volume, ITV: internal target volume, IGTV: internal gross tumour volume, PTV: planning target volume, Tx: prescription dose, 4DCT: 4-dimensional computed tomography, IGRT: image-guided radiotherapy, fx: fraction, MU: monitoring unit.}
\small
\begin{tabular}{p{0.16\linewidth} p{0.25\linewidth} p{0.25\linewidth} p{0.25\linewidth}}
	\hline 
	        Institution & \multicolumn{1}{c}{MUHC} & \multicolumn{2}{c}{CHUM} 	 \\
		Technique & \multicolumn{1}{c}{3D-CRT} & \multicolumn{1}{c}{VMAT} & \multicolumn{1}{c}{CyberKnife} \\
		\hline \\
		Dose prescription  & Dose normalized to 100 \% at Tx, 95\% of PTV receives Tx or higher (D95\% $\geq$ Tx) &  \multicolumn{2}{>{\centering\arraybackslash}p{0.5\linewidth}}{Dose normalized to 100 \% at Tx which covers 95\% or more of the PTV} \\
		Dose planning procedure/ \linebreak calculation algorithm & Forward planning using Eclipse (Varian, USA)/ superposition-convolution algorithm with heterogeneity correction & Inverse planning with RapidArc (Varian, USA)/ superposition-convolution algorithm with heterogeneity correction & Inverse planning with Multiplan (Accuray, USA) / Monte Carlo calculation \\
		
		Beam type & 6 MV photon & 6 MV photon & 6 MV photon \\
		
		Target volume definition & ITV: drawn from 4DCT using maximum intensity projection \newline PTV: ITV + 5 mm margin & IGTV: drawn on extreme phases of 4DCT to represent its full extent \newline PTV: IGTV + 5 mm margin & GTV: drawn on breath hold, corrected if needed for deformation/rotation using extreme phases \newline PTV: GTV + 5 mm margin \\ 
		
		Dose fractionation & 50 Gy in 5 fx: tumor at central location and/or close to critical organs (chest wall/large vessels/spinal cord) \newline 34 Gy in 1 fx: otherwise, upon patients' request for shorter treatment \newline 48 Gy in 3 fx: otherwise & \multicolumn{2}{>{\centering\arraybackslash}p{0.5\linewidth}}{50 Gy in 5 fx: tumour at central location \newline 60 Gy in 5 fx: peripheral tumour close to OARs \newline 60 Gy in 3 fx: otherwise} \\
		
		Dose constraints to OARs & 50 Gy in 5 fx: RTOG 0915  \newline 48 Gy in 3 fx: RTOG 0915 \newline 34 Gy in 1 fx: RTOG 0813 & \multicolumn{2}{>{\centering\arraybackslash}p{0.5\linewidth}}{Timmerman et al. \cite{Timmerman08} } \\

		Immobilization & BodyFix (Elekta Oncology, Norcross, GA) & BodyFix (Elekta Oncology) & Vac-Lok (Civco Medical Solutions, Orange City, IA) \\
		IGRT & CBCT at every fraction & Pre- and mid-treatment CBCT at every fraction & Real-time target tracking \\
		Plan verification & Independent MU check & Independent MU check, daily dynalog verification & Independent MU check \\ 	
		
  	\hline
\end{tabular}
\label{tab:TPdetails_train}
\end{center}
\end{table}
\end{landscape}
\restoregeometry

\begin{table}
\begin{center}
\caption{Detailed radiotherapy procedures used for the validation cohort.}
\begin{tabular}{p{0.3\linewidth} p{0.6\linewidth}}
\hline 
Institution & \multicolumn{1}{c}{WashU} \\
Technique & \multicolumn{1}{c}{3D-CRT} \\
\hline
Dose prescription & Dose generally prescribed to 80\% isodose line (range 60-90\%) and covers >95\% of PTV  \\
Dose planning procedure/ \linebreak calculation algorithm & Forward planning with 7-11 non-coplanar beams using Pinnacle (Philips, Netherlands)/ superposition-convolution algorithm with heterogeneity corrections \\
Beam type & 6 MV photons \\
Target volume definition & ITV: drawn from 4DCT using maximum intensity projection \newline PTV: ITV + 5 mm margin\\
Dose fractionation & 50-60 Gy in 5 fx: central location or close to critical organs \newline 54 Gy in 3 fx: all others \\
Dose constraints to OARs & 50-60 Gy in 5 fx: RTOG 0813 \newline 54 Gy in 3 fx: RTOG 0618  \\
Immobilization & Abdominal compression (CDR systems, Canada) \\
IGRT & CBCT at every fraction with KV fluoroscopy \\
Plan verification & Independent MU check \\
\hline
\end{tabular}
\label{tab:TPdetails_val}
\end{center}
\end{table}

\begin{table}
\begin{center}
\caption{Odds ratios of candidate variables, bin boundary used for discretization, and frequency of selection obtained by bootstrapping the KS variable filtering. P-values were adjusted for multiple comparison using a method by Benjamini and Hochberg \cite{Benjamini95}. *variables selected for the BN modeling stage. $\dagger$ taken as a percentage change from baseline.}
\begin{tabular}{ l c c c }
	\hline
	& Odds ratio (p-value) & Bin boundary & Selection frequency  \\ \hline
	Biological variables &  & &  \\
	\quad OPN (baseline)* & 0.887 (0.886) & 54.2 ng/ml & 0.394 \\
	\quad OPN (6 weeks$\dagger$) & 1.150 (0.886) & 80.9 \% & 0.133 \\
	\quad IL8 (baseline) & 2.862 (0.210) & 31.0 pg/ml & 0.228  \\
	\quad IL8 (6 weeks) &  0.404 (0.637) & -60.4 \% & 0.264 \\
	\quad ACE (baseline) & 1.999 (0.529) & 141.1 ng/ml & 0.308\\
	\quad ACE (6 weeks)* & 0.002 (0.010) & -15.8 \% & 0.782 \\
	\quad IL6 (baseline) & 0.070 (0.657) & 7.0 pg/ml & 0.2 \\
	\quad IL6 (6 weeks) & 1.106 (0.886) & -7.0 (\%) & 0.058\\
	\quad a2M (baseline) & 0.553 (0.638) & 5.3 mg/ml & 0.328 \\
	\quad a2M (6 weeks) & 0.848 (0.886) & -7.6 \% & 0.142 \\
	\quad TGFb (baseline)* & 1.866 (0.540) & 42.2 ng/ml & 0.504\\
	\quad TGFb (6 weeks) & 0.493 (0.610) & 1.4 \% & 0.053 \\
	Dosimetric variables & \\
	\quad MLD (ipsilateral) & 2.400 (0.391) & 13.8 Gy & 0.107 \\
	\quad V5 (ipsilateral)* & 5.685 (0.060) & 42.4 \% & 0.454 \\
	\quad V105\%* & 5.848 (0.023) & 1.4 cc & 0.668 \\
	\quad Fraction size & 0.752 (0.886) & 20 Gy per fraction &  0.142 \\
	\quad PTV volume & 1.932 (0.518) & 20.5 cc & 0.064 \\
	\quad MHD & 1.945 (0.529) & 9.0 Gy & 0.153 \\
	Clinical variables &  & \\
	\quad PTVCOMSI*& 0.379 (0.391) & 0.5 & 0.448 \\
	\quad Age & 1.172 (0.886)  & 69 & 0.121 \\
	\quad Smoking & 1.077 (0.945) &  & 0.146 \\
	\quad IP & 1.300 (0.768) & & 0.061 \\
	\quad Central tumour & 1.800 (0.854) & & 0.068 \\
	\quad COPD & 0.750 (0.886) & & 0.120 \\
	\quad ACE inhibitor & 0.800 (0.638) & & 0.054 \\
	\hline 
\end{tabular}
\label{tab:variables}
\end{center}
\end{table}

\end{document}